Forthcoming in Synthese

## Laws of Nature and the Reality of the Wave Function[1]

Mauro Dorato
Department of Philosophy, Communication and Media Studies
Via Ostiense 234, 00146, Rome, Italy

Key words: wave function; realism; nomic realism; dispositionalism; configuration space realism

One of the words of the philosophical jargon that generate more incomprehension between physicists and philosophers of physics is "realism". Even when theoretical physicists are sensitive to the conceptual problems of quantum mechanics – for instance by recognizing that the lack of a quantum theory of measurement is a serious problem – they typically display indifference about the problem of determining the ontological status of the wave function.[2] On the contrary, in the philosophical debates the attempt to understand its role in the interpretation of quantum mechanics is central. Why is there this difference between the two communities?

The physicists' skeptical attitude depends, at least in part, on their instinctive mistrust of philosophical debates concerning "-isms", especially when they are applied in a generic way to all scientific knowledge. Furthermore, physicists are usually not trained to ask themselves why the algorithms that they use work as efficiently as they do, so that the "shut up and calculate" attitude prevails. This fact creates the conviction that the predictive use of mathematical models of phenomena (Schrödinger's equation included) exhausts their meaning, so that their apparent and explicit agnosticism about the status of $\Psi$ often borders with an unreflectively defended, implicit form of instrumentalism. In the case of quantum theory, the physicists' approach has been reinforced by well-known historical and sociological reasons that here are superfluous to recall.[3] As a consequence, significant connections between the often-acknowledged importance of the measurement problem and related issues about the nature of the wave function escape them.

On the contrary, philosophers of quantum mechanics have correctly stressed the importance of such connections, but have defended different opinions on the ontological status the wave function $\Psi$: is it a bookkeeping device or it denotes an *entity* of some sort? And, in the latter alternative, is it a *mathematical* or a *physical* entity? And if it is physical, is it primitive and fundamental or derived and emergent?

---

[1] Thanks to Albert Solé and to two other anonymous referee for their comments and criticism.
[2] Of course, this neglect is compatible with a natural ontological attitude towards endorsing the existence of elementary particles (Fine 1984).
[3] Born's original attitude towards his own rule explicitly mentioned probability of measurement outcomes, a claim that in turn depended on Bohr's view that in quantum physics measurement is a primitive and fundamental notion. See Cushing (1994).

In trying to answer these questions, philosophically sophisticated instrumentalists about the wave function like the physicist Carlo Rovelli follow Bohr's lead, with important corrections coming from relationalism, information theory and subjective philosophies of probability.[4] On the other side of the camp, we find philosophers defending the reality of $\Psi$, often inspired by the following quotation, taken from John Bell's approach to Bohmian mechanics: "no one can understand this theory until he is willing to think of ψ as a real objective field rather than just a 'probability amplitude.' Even though it propagates not in 3-space but in 3N-space" (1987, p. 128).

In this paper I will try to tackle the above questions by critically evaluating three different "realistic" (anti-instrumentalistic) positions on the wave function. In assessing their merit, I will regard as essential their capacity to account for the world of our experience in space and time:

(1) Realism about laws or *nomic realism*, a position to which the ontologically unclear view of $\Psi$ as a "nomological *entity*" (Dürr Goldstein's and Zanghì 1997, Goldstein and Zanghì 2013) seems to be committed.[5] I will regard nomic realism as just *sufficient* for the reality of the wave function, in such a way that Humean accounts of $\Psi$ (Callender 2014) – that here I will not discuss – are *not* ruled out. This first argument in favor of wave-function realism therefore amounts to claiming that *if* fundamental laws exist – whatever this means – then entities like $\Psi$, which enter in one of them as an essential component, should also exist.

(2) Dispositionalism (Dorato and Esfeld 2010, Esfeld et. al 2013), which is really a form of "indirect realism" about $\Psi$, in the sense suggested by Monton: "The wave function doesn't exist on its own, but it corresponds to a property possessed by the system of all the particles in the universe" (Monton 2006, p. 779). Dispositionalism clarifies what kind of property the wave function regarded as an mathematical entity *denotes*, and is realistic about it to the extent that the quantum state of the universe is identified with such a property.

(3) Configuration and wave function-space realism (Albert 1996, 2013, North (2013).[6]

I will conclude that all of these views are committed to consider $\Psi$ to be a mathematical entity,[7] despite the different ways in which they connect their respective ontological claims with spatiotemporally extended entities, namely with the view that quantum physics requires a primitive ontology of entities located in spacetime (Dürr et al 1992, Allori et al. 2008, Allori 2013). As we will see, the fact that the first two views are ontologically on a par *vis à vis* the status of $\Psi$ is a consequence of quantum holism, a feature of quantum theory that bring them closer to each other. The third position, on the contrary, will be shown to be a merely speculative attempt to derive such a primitive ontology from a reified mathematical posit. I will therefore conclude that there remain only two available alternatives about the status of $\Psi$:

(i)     Instrumentalism as a consequence of the implausibility of the three positions above: $\Psi$ has an empty reference;

(ii)    The claim that whatever is denoted by the wave function exists abstractly and not in spacetime;

---

[4] For an assessment of these latter elements, see Dorato (2013). For a general exposition of Rovelli's philosophy of quantum theory, see Rovelli and Laudisa (2013).

[5] While these authors may be interpreted as being neutral with respect to the question whether laws are *its* or *bits* (to use Callender's (2014) felicitous expression), I will argue that in their texts they seem to defend the former option.

[6] An informative case-study *vis à vis* these options is in Belot (2012). Many-worlds realism about $\Psi$ will be discussed within the view of configuration space realism (Vaidman 2012).

[7] For an account of "abstract", see next page.

Two remarks are appropriate at this point.

1) The first is that in order to clarify the implications of (ii), it is essential to clarify the notion of "abstract". This task, however, is tricky, not only because there is no agreement about how to go about it, but also because the "definitions" are typically given by "way of negation" (see Rosen 2014). Here I will take for granted a characterization that is widely shared also in the philosophy of mathematics, namely that *an entity X is abstract if and only if it is either not in spacetime or is causally inert or both*. Despite the difficulties noted by Rosen, the wider scope afforded by a disjunctive formulation and its connection with the philosophy of mathematics will prove rather useful for my purposes. In fact, in what follows I will assume, as it is standard to do, that if mathematical entities exist, they are abstract because they are *neither* causally efficacious *nor* in spacetime. Assume furthermore, as is plausible to do, that an entity $X$ is physical if and only if it is *either* in spacetime *or* causally active or both. Since if they exist mathematical entities are *neither* in spacetime nor causally active, it follows that an entity is mathematical if and only if it is not physical.

It is important to note that the disjunctive definitions given above allows for the possibility that an entity could be abstract because non-spatiotemporal and yet be physical (non-mathematical) because causally active. Agreed: various approaches to causation imply that an entity $X$ can be causally active if and only if it is in spacetime. However these a priori theories of causation should not detain us from considering the possibility that some *physical* entity could be *abstract*, as required not only by Albert (2013)'s view of the ontological status of the wave function, but also by current attempts to "derive" spacetime from a more fundamental physical entity.

2) While here I will not try to adjudicate between (i) and (ii), it should be already clear to the reader why instrumentalism about the wave function is not as unmotivated as it has seemed to be to many philosophers, in particular to those inclined towards the rejection of abstract entities.

The dilemma generated by (i) and (ii) might seem unfair to a recent proposal by Maudlin, who claims that the wave function denotes a *wholly new physical entity*, one that does not belong to any of the available metaphysical categories, "abstract", "concrete" and "causally efficacious" included (Maudlin 2013). The reason for my dismissing this proposal from the very outset is that at the current state of our knowledge it is wholly unjustified, since it is equivalent to saying that *we don't understand what the wave function is*. This implies that so far we cannot give any account of how it relates to macroscopic objects living in three-dimensional world or events spread out in four-dimensional spacetime. Lacking such an explanatory connection, the thesis that the wave function exists as a new physical entity is a mere promissory notice, and as such does not take us very far. Consequently, on the basis of the requirement expressed above, here I can afford to ignore it.

The plan of the paper is as follows. In the next section I will begin by offering some arguments in favor of the importance of embracing a piecemeal and pluralistic perspective on scientific realism, so as to respond to the physicists' above-mentioned criticism about the emptiness of the instrumentalism/realism debate when applied to *all* physical theories. This section has the main aim of going some way toward justifying why here I focus on a single component of a particular physical theory, namely an entity figuring in a fundamental dynamical law. In the third section, I will critically discuss Goldstein and Zanghì's realism about the wave function, by them regarded as "a nomological entity" (Goldstein and Zanghì 2013), in order to show that their position seems to be committed to a sort of primitivism about laws defended by Maudlin (2007). As anticipated above, this interpretive move does not rule out the compatibility between Humeanism and nomological view of the wave function. In the fourth section I will discuss the recent surge of dispositionalism about quantum mechanics – variously defended by Suárez (2004, 2007), Dorato (2007), Dorato and Esfeld (2010) and Esfeld et al. (2013). In the fifth I will criticize the reifying attitude of configuration space realism (Albert 1996, 2013, North 2013). In the final section I will draw the moral suggested by the previous sections by concluding that realists about $\Psi$ are committed to the view that it denotes an abstract but non-physical, and therefore a purely mathematical, entity in the sense specified above. Consequently, supposing that an ontology of mathematical or even just abstract entities needs an instantiation in spacetime by non abstract entities, whatever their nature is, the latter will have to be regarded are the truth-makers of Schrödinger equation. Alternatively, a rejection of abstract entities – that here cannot be motivated – will force one to endorse a form of instrumentalism *about ψ.*

In what follows, and also for lack of space, I will focus mostly on nomic realism, since dispositionalism and configuration space realism have received more attention in the literature (the latter view, in particular, has been the subject of a recent book edited by A. Ney and D. Albert 2013). Furthermore, since the wave function figures in a law, in order to answer questions about its status it is important to understand what it means to claim that it is a *nomic entity*. As anticipated above, if laws existed as primitive ontological posits featuring in all physically possible worlds, realism about the wave function would follow, a plausible inference that entails a privileged attention to this position. Finally, nomic realism shares with dispositionalism the view that modal features of the world exist *de re* and are not a mere property of our language and models (as they are in Humeanism). Therefore, it is important to start discussing one of the two views and see to what extent it differs from the other, in particular *a propos* of the problem of the quantum state (see Dorato and Esfeld 2014). The choice to devote more space to nomic realism than to dispositionalism is also dictated by the fact that the former view is more difficult to clarify and is therefore in need of further analysis.

## 2 Piecemeal Realism and Selective Instrumentalism[8]

The complaint of the working physicists about scientific realism in general is justified to the extent that "realism" and "instrumentalism" are subject to ideological discussions that are often quite removed from the practice of physics and the content of single scientific theories. I shall suppose that being a realist about physical theories is the consequence of a *stance* (van Fraassen 2002), that is, of a general, overarching attitude toward the *aims* of physics.[9] As it happens with epistemic as well as non-epistemic aims, there is no *a priori* assurance that they can be always realized. In our case, it is not wholly absurd to claim that one could be instrumentalist about physical theory *T* and realist about theory *T'*, according to the kind of evidence (and other epistemic virtues) that *T* and *T'* can provide. There is historical evidence that this "piecemeal" or "contextual" approach to realism (McMullin 1984, Miller 1987, Fine 1991, p. 87) might be justified even for different posits of a *single* physical theory: Lange (2002) shows how in classical electromagnetism one ought to be realist about the electromagnetic field but instrumentalist about Faraday's lines of force.[10] Couldn't the same "pluralistic view" hold also for our case, namely couldn't one be instrumentalist about the wave function while granting existence to quantum entities like particles and fields (or whatever is more fundamental between the two)?

---

[8] "Selective anti-realism" is the title of a paper by McMullin (1991)

[9] This aim involves the commitment both to the claim that well-confirmed physical theories are at least approximately true given the evidence (the epistemological claim) and to the claim that they purport to describe a mind-independent, non-directly observable world (the metaphysical claim). These claims are obviously related.

[10] For an early defense of piecemeal realism, see Miller (1987).

Philosophical views that go a long way toward justifying piecemeal or contextual realism within a single theory are various and numerous. Since they are well known, they don't need to be discussed here. For instance, by denying any realist commitment to theoretical laws while granting existence to theoretical entities, *entity realism* (Hacking 1983, Cartwright 1983) is perhaps the best known of these views. Also Chakravarrty's (2007) distinction between detecting and *auxiliary* properties, which is meant to distinguish those components of the theory that deserve a realistic commitment from those that don't, points to a related kind of selective realism. Even epistemic structural realism, by separating relational or structural properties from intrinsic or non-structural properties, could be regarded as a form of contextual realism.

Since all of these views provide good initial reasons to believe in some kind of contextual realism even within the same theory, the debate about the status of the wave function can be considered as an additional case study to evaluate its plausibility. In particular, it becomes conceivable to ask whether one can maintain the view that quantum mechanics is about some beables in spacetime, without having to accept the reality of the wave function. Also the converse claim becomes conceptually possible and worth investigating: the wave function could be all there is, and the world of three-dimensional objects in space, both microscopic and macroscopic, could be only an appearance or, at best, "emergent" (Vaidman 2012, Wallace 2012).

In a word, piecemeal realism entails that a realistic attitude about the wave function is not the only way to defend some form of "realism" about quantum mechanics. One could support a "flash" or a "density-of-stuff" ontology (two variants of GRW), or an ontology of worldlines traced by particles endowed with well-defined positions (as in Bohmian mechanics), as *primitive* ontologies for observer-independent formulations of quantum mechanics (Allori, et. al. 2008), and be at the same time wholly instrumentalist about the wave function. Recall that "primitive ontologies", if they entail a commitment to something concretely existing in spacetime (see also Allori 2013), might be naturally regarded also as the ground for less fundamental, macroscopic entities. However, since, as we are about to see, the wave function could be primitive in the sense of being fundamental or irreducible, two questions arise:

(Q1) can we claim that primitive ontologies invoking entities in spacetime do not require any sort of wave function realism or should we attribute the latter only a less fundamental ontic status?

(Q2) conversely, is it legitimate to regard the wave function as ontically fundamental and therefore consider spatiotemporally extended entities as emergent or even illusory?

Before entering in *medias res*, let me add three brief remarks.

1. Note that "primitive" can used in two different senses: in order to avoid ambiguities, let us stipulate that "primitive ontology" refers to entities in spacetime, while "primitive" by itself just refers to what is ontically fundamental, even if not spatiotemporally extended.

2. Since the ontological posits of physical theories cannot wildly contradict that manifest image of the world from which they gain all their evidential support, they'd better help us to specify, at least in principle, how to recover the latter from the former. While we cannot assume that spacetime is not emergent from something else as quantum gravity seems to suggest (Butterfield and Isham, 1999, Rovelli 2004, ch.10, Kiefer 2007, Lam and Esfeld 2013), the explanatory task of giving an account of the world of our experience is certainly harder if primitive ontologies in the sense of Allori et al. (2008) are themselves emergent from something that is not spatiotemporally extended, as hypothesized by Q2.

3. Finally, note that if the philosophers of physics' task consists in examining the credentials of realism only for a particular component of the mathematical model of quantum mechanics (namely the wave function), the specificity and focus of the inquiry should be much more attractive for physicists; the recent interest for the claim that the quantum state does not reduce to information (Pusey, Barrett, and Rudolph 2012, Schlosshauer and Fine 2012) is evidence for this more "pragmatic" attitude. In a word, an additional advantage of this selective form of realism is therefore the abandonment of generic defenses of scientific realism as such; consequently, the above-mentioned gap between physicists and philosophers of physics can, at least to a certain extent, be bridged.

## 3. The wave function as a nomological entity: the first way of reading wave-function realism

In Goldstein and Zanghì (2013, p.92), the wave function is defined as "nomological", that is, as something that, while relating to laws of nature, has nevertheless some sort of ontological significance (they use the expression "nomological *entity*"). The next two quotations somewhat clarify what they have in mind: "The fact that Bohmian mechanics requires that one takes such an unfamiliar *entity* seriously bothers some people. It does not bother us that much…" (Goldstein and Zanghì 2013, p.93, italics added). More in detail, the wave function according to Goldstein and Zanghì is neither *everything* (as in Everettian quantum mechanics) nor *nothing* (as in purely instrumentalist views), but "it is something"… "The wave function would seem part of the ontology. It is real in that sense. It's not subjective in Bohmian mechanics − it has a rather real role to play: it has to *govern* to motion of the particles." (Goldstein and Zanghì 2013, p. 92, italics added).

In their view, however, whatever governs needs something to be governed, namely the entities that constitute the *primitive* ontology of the theory: in the case of Bohmian mechanics, this ontology is given by particles endowed with a definite position in spacetime. Following Allori et. al. (2008), we can generalize this notion of primitive ontology also to other theories alternative to standard QM and widely different from Bohmian mechanics, namely those that are proposed by dynamical collapse models. In the two versions of GRW ("flashes" and "density of stuff") are the primitive ontological assumptions, since both flashes and fields are in spacetime (see Allori et al. 2008). All these "heretical" theories share a common structure, since they accord ontic primacy *to stuff in spacetime* – while the role of the wave function is to govern the behavior of matter.

However, what does it mean to claim that the "wave function is real" and non-subjective because "it governs the motion of particles", as the last quotation affirms? As a partial answer to this question, Goldstein and Zanghì specify that some kind of ontic role for the wave function be suggested by Bohm's "guiding" equation, according to which the velocity of each of the $N$ particles of the universe is a non-local function, via the universal $\Psi$, of the positions of all the other particles. This means that $\Psi$ determines the speed and the direction of motion of each particle as a function of the *global* configuration of all the other particles in the universe. Consequently, the real nomological entity is, properly speaking, only the wave function of the universe $\Psi$. The universe is "the only genuine Bohmian system" (Goldstein and Zanghì, 2013, p. 94), since the wave function of a subsystem $\psi$ becomes definable only in terms of the wave function of the universe $\Psi$ and the whole set of configuration of the remaining particles. Given the non-local character of the theory, this is only to be expected, even though for all practical applications what one deals with in Bohmian mechanics are *subsystems* of the universe as a whole. Since the wave function $\Psi$ is a global feature of the universe (the wave function of the universe), it appears even rather reasonable to consider it a *law*, since laws are valid always and everywhere.[11]

Unfortunately, these plausible remarks, while strong enough to put $\Psi$ in the nomological category, still do not suffice to answer the crucial, italicized question raised in the previous paragraph. If one starts from the assumptions that $\Psi$ is not unreal but also not ontologically primitive, the claim that it is a "lawlike entity" is rather natural. However, until we are told what it means to claim that physical laws are not part of the primitive ontology of a theory, and yet are to be regarded as entities, labeling the universal wave function as a "nomological entity" is a purely *verbal* solution to our problem.

---

[11] It could be objected that in the absence of a Bohmian theory of quantum gravity, it is meaningless to attribute a wave function to the universe. This objection however, would be unfair, since here we are trying to fathom the metaphysical consequences of non-relativistic Bohmian mechanics, so that the arena that we need to presuppose to formulate the theory is Newtonian spacetime (or for relativistic extensions, Minkowski spacetime with the addition of a privileged foliation).

Confusion on this point possibly explains the following perplexing passage, which seems to contradict Goldstein and Zanghì's implicitly realistic stance about the universal wave function, previously referred to as a nomological "something": "we have never heard anyone complaining about classical mechanics because it involves a weird field on phase space and asking about what kind of thing that is. No one has any problem with that. Everyone knows that *the Hamiltonian is just a convenient device in terms of which the equation of motion can be nicely expressed*" (Goldstein and Zanghì 2013, p. 98, my italics). By analogy, the italicized sentence in this quotation leads us to think that also the universal wave function is not an *entity*, as our authors claim, but only a "convenient [mathematical?] device"[12]; however, in what sense is their view of the wave function not fully instrumentalistic, or more precisely, Humean?

Independently of hermeneutical exercises, which here are out of place, in order to clarify what a realistic nomological understanding of the universal wave function is committed to, one should *first of all* accept the following point due to Maudlin: the wave function is, strictly speaking, a *mathematical function* from configuration space to complex numbers (Maudlin 2013, p.129). Note that here "abstract" has the same meaning that has been clarified before. Therefore, the real question is whether such a function ─ considering the role it plays in the guiding equation ─ refers to "something" physically real but not primitively real (i.e. not in spacetime), as Goldstein and Zanghì have it, and what the nature of this "something" is (for instance, causal, if "abstract" does not entail causally inert, but only non spatiotemporal).[13] We may end up concluding that also this "something" is *also* causally inert and therefore purely mathematical, of course, and this is exactly what I will show in the remainder of this section. In any case, in order to avoid misunderstandings, from now on the term "universal wave function" will refer to this (possibly existing) "something".

*Secondly*, it must be admitted that the kind of ontic dualism created by the distinction between particles and laws need not imply antirealism for one of the two "entities": both can be regarded as real. However, any form of ontological dualism creates the problem of the *relation* between *two* types of entities,[14] one of which (particles), in the option we are considering, is ontologically and epistemically prior.

One way to understand this relation in our context is to read it in terms of ontic grounding: primitive entities (and Bohmian particles among them) can be treated as more fundamental with respect to $\Psi$ for at least two reasons. The first, which is one of the strengths of Bohmian mechanics, is that, by existing in spacetime, particles have an ontic as well as an epistemic priority because they account for the properties and the existence of the three-dimensional objects of our experience. Such objects can in fact be regarded as partially *constituted* by these particles, as in the part-whole relation. The other, more controversial sense in which particles are more fundamental than laws lies in the hypothesis that particles *exemplify* or *instantiate* relations holding among the various stages of a temporal process governed by Bohm's guiding equation (see Esfeld et. 2013).[15] Such relations would not exist without the *relata* (the particles): how could laws "govern" if they are not instantiated by something?

---

[12] The word in square parentheses is my addition.
[13] Note once again that the disjunctive definition of abstract given in the previous section allows for such a possibility.
[14] Also Callender (2014) raises this question for the relation between Bohmian particles and the guiding-law, when the latter is regarded as an "it" rather than a "bit".

Now we are in a better position to understand why, formulated in this way, realism about the universal wave function calls into play the philosophy of laws of nature,[16] which, in the anti-Humean camp, is divided between nomic primitivism (Maudlin 2007) and dispositionalism (Mumford 2004). Briefly put, nomic primitivism is the view that laws exist and are ontologically prior with respect to properties or causal powers, in the sense that they fix which of them can be attributed to physical entities. Dispositionalism is the view that properties or causal powers are ontologically primary with respect to laws, since the latter presuppose the former as their truth-makers. In the rest of this section I will discuss the first option, which is more suitable to attribute a direct and fundamental ontic role to the universal wave function, while in the next section I will discuss dispositionalism.

On the basis of the fact that the physical realm need not include just what is in spacetime (and is therefore part of the primitive ontology) there seem to be only three alternative ways of interpreting the claim that the universal wave function $\Psi$ is a nomological *entity*:

(i)     Either there is no principled distinction between physical and mathematical entities, so that $\Psi$ is *both* a physical and an mathematical entity; or, if such a distinction is clear-cut

(ii)    $\Psi$, regarded as a nomological entity is abstract and *non*-physical, in any sense of physical, so that the abstract mathematical function occurring in Schrödinger's equation *refers* to something (the quantum state) that is also purely mathematical;

(iii)   $\Psi$ is a nomological and physical entity.

Let me discuss these three hypotheses in turn.

(i) The view that in physical theories *in general* there is no clear demarcation between mathematical and physical ontological commitments cannot be correct, despite claims that there is no *epistemic* way to distinguish them (Ladyman and Ross 2007 and Ladyman 2014). So it cannot be the case that $\Psi$ is *both* a physical and mathematical entity. Four briefly stated objections to this hypothesis suffice to refute it.

---

[15] This reading was applied by Dorato and Esfeld (2010) to GRW-type theories.
[16] See Dorato (2005).

First of all, mathematical hypotheses are not decided on experiments, as physical conjectures are, so there seems to be a clear-cut distinction between the epistemology of mathematics and that of physics; if this is the case, there is an important epistemic way to distinguish ontological claims suggested by mathematics from those suggested by physical theories.

Secondly, mathematics is not a safe guide to ontology, for the plausible reasons already illustrated by Maudlin (2013) and that here will be taken for granted. However, if there were no distinction between physical and mathematical ontological claims as (i) suggests, the same conclusion would apply to all physical theories. In this case, also such theories would become completely useless to evaluate ontological claims suggested by their formalism and by experimental results. This skeptical conclusion would apply to all of their posits, and therefore also to $\Psi$. Neither realism nor antirealism about the universal wave function could be justified, and the main question of the paper would remain unanswered. In any case, this agnosticism cannot support any form of wave function realism.

Thirdly, if mathematical entities denote, and nominalism is wrong, they typically refer to an abstract realm of causally inert, non-spatiotemporally extended entities (see the definition given above). However, if there were no sharp demarcation between mathematical and physical ontology, then also the latter would have to be regarded as populated by purely abstract and causally inert stuff. But then, how could we explain the world of our experience – which *appears* to be causally active and spatiotemporally extended – by presupposing just an ontology of unobservable, abstract and causally inert entities? There must be something causal or spatiotemporally extended that *instantiates* this abstract structure, a requirement that (i) cannot satisfy even in principle.

Fourthly, suppose nevertheless that physical and mathematical structure are one and the same. Mathematical structures are identified up to *isomorphism*s; therefore, if (i) is correct, also the ontology of physical theories must be subject to the same constraint. However, this fact has unpalatable consequences that are quite independent of wave function realism. Van Fraassen has formulated a strong objection against a collapse of physical onto mathematical structure, which also applies to (i): "*what has looked like the structure* of something with unknown qualitative features *is actually all there is to nature*. But with this, the contrast between structure and what is not structure has disappeared. Thus, from the point of view of one who adopts this position, any difference between it and 'ordinary' scientific realism also disappears. It seems then that, *once adopted*, it is not to be called structuralism at all! For if there is no non-structure, there is no structure either." (2006, pp. 292-293). It then follows that ordinary realism about three-dimensional physical objects in spacetime must be endorsed, against the idea that physical ontology can be identified only up to isomorphism. With these objections, we can regard (i) as rejected.

ii) Suppose now that, granting a clear demarcation between physical and mathematical ontology, the wave function is itself a *mathematical* entity denoted by the mathematical function occurring in Schrödinger's equation. Given the definition of "abstract" given above, $\Psi$ is neither causally active nor spatiotemporal. In this sense, it is a non-physical but nomological entity, evolving in configuration space, but *referring* in some sense to the physical world.[17] Notice that this hypothesis is *not* equivalent to Albert (1996)'s claim that configuration space is the fundamental *physical space* from which three-dimensional space emerges. According to Albert's desiderata in fact, to be briefly discussed and then rejected in section 5, configuration space is physical (non-mathematical) despite its being abstract (non-spatiotemporal).

The hypothesis (ii) that we are now discussing is not as absurd as it might seem *prima facie*. For example, it could be backed up by Psillos' claim that physical theories (quantum mechanics included) are *directly* about mathematical models, to which a realist, literalist interpretation of physical theories is committed (Psillos 2011) and refer only *indirectly* to the physical world (Morgan and Morrison 1999), *via* the mediation of mathematical models essentially defined by laws of nature (Giere 1988). In this sense, the universal $\Psi$ could be regarded with no contradiction as a "nomological, but mathematically abstract, non-physical *entity*", a conclusion that is indeed the most reasonable of the three options mentioned above. As such, it is an important piece of evidence in favor of the main thesis of the paper. In order to clarify its consequences and avoid possible objections, let me add the following three points.

1) Given the radical separation between physical and mathematical entities that is presupposed in (ii), Psillos' claim should be interpreted as entailing that the ontology of a physical theory *includes* also mathematical models of physical phenomena. However, these models are strictly speaking non-physical, and are "physical" only to the extent that they *refer* to physical entities that are either causally active or spatiotemporally extended or both.

2) If the universal wave function is to be regarded as mathematical but real because quantum mechanics is directly about it, it will also have to be regarded as *causally inert*. Consequently, the view that the wave function can *affect* in some way the motion of the particles needs to be abandoned. This task would entail that expressions that are frequently encountered in the Bohmian literature – $\Psi$ is something that literally "governs" the motion of particles by pushing them around – are mere metaphors devoid of any physical meaning. Particles are governed in the sense that they obey the laws of Bohmian mechanics. Even though I will not purse this line of argument here, one could argue in favor of the causal inertness of the wave function by insisting on the fact that the validity of the action/reaction Principle requires the existence of a genuine causal interaction between any two entities.[18] I addition, the fact that the wave function lives in configuration space while the particles are in three space renders an account of the casual influence of the former over the latter quite difficult to understand.

---

[17] How this reference should be understood will be discussed below.

3) If the universal wave function is a mathematical entity, it becomes even more urgent to recognize, along with Allori et al. (2008), that there is some primitive ontological posit that quantum theory is *also* and primarily about (namely particles): this seems the only way to recover the world of our experience, which is in spacetime.

In a word, in order to formulate in a correct way the ontology of Bohmian mechanics and that of the GRW-type theories, according to the hypothesis (ii) we need both primitive entities in spacetime and a nomological but mathematical entity like the universal wave function.

We can now turn to the last of the three hypotheses listed above, which maintains that the universal wave function is a nomological but *physical* entity. We will see that the difficulties of this position force us to choose the view that $\Psi$ is a mathematical entity as in the hypothesis ii) just examined.

(iii) If the wave function is viewed as some sort of causally inert but *physical* and nomological "blob" (see French 2013), it can either be regarded as spatiotemporally extended[19] or as a primitive physical fact about any physically possible world, in the sense advocated by Maudlin's ontic nomic primitivism (2007). Here I will focus just on these two alternatives, while in the next section I will discuss dispositionalism.[20]

Goldstein and Zanghi's (2013)'s claim that the wave function is not a physical entity in the sense advocated by the former alternative seems correct: it is only if there only one particle in the universe that the wave function could be regarded as a spatiotemporally extended entity. Unlike spatiotemporally extended stuff – particles, flashes, density of matter – the universal wave function of $N$ particles cannot be characterized as causally active, for the reasons already explained in (ii). So, by being neither spatiotemporal nor causally active, it is non-physical.

However, couldn't it be the case that each particle has it own wave function guiding its motion,[21] so that the universal wave function is a sort of wavelike "pattern" of spatiotemporally *local* matters of facts? An immediate, and I think fatal objection to this proposal, is entanglement: any sort of "pattern" supervening on the Humean mosaic has to be global and cosmically extended.

---

[18] As is well known, in Bohmian mechanics the particles cannot back-react on the field.

[19] Recall that the hypothesis that $\Psi$ is causally active has been rejected before.
[20] Recall that the view that the universal wave function lives in a $3N$ dimensional, physical configuration space (Albert 1996) will be discussed in a later section.
[21] This view is explored but not fully endorsed by Norsen (2010).

If the wave function of any particle $x$ is entangled with that of any other particle $y$, the physical "blob" of the wave function of the universe ends up being defined in a $3N$ configuration space. Recall that the wave function of any subsystem is used only FAPP (**F**or **A**ll **P**ractical **P**urposes), the real wave function being that of the universe (see above). This first alternative can therefore be rejected.

Let us now turn to the second alternative, according to which the nomological, physical reality of the universal wave function follows as a consequence of nomic primitivism. Nomic primitivism can be understood in a conceptual or in an ontological way. The conceptual approach argues that the *concept* of laws is irreducible to any other related *concepts* (causation, counterfactual, property, disposition, etc.) The ontic version is committed to realism about laws of nature, thereby transforming conceptual nomic priority into ontic nomic priority, and using the former to defend the latter. Here is a particularly clear formulation of ontic nomic primitivism: "If there were no laws, then there would be no causation, there would be no dispositions, there would be no true (nontrivial) counterfactual conditionals. By the same token, if there were no laws of nature, there would be no perception, no actions, no persistence. There wouldn't be any tables, no red things, no things of value, not even any physical object." (Carroll 2004, p. 10).

Maudlin defends an ontic understanding of the priority of laws more explicitly than Carroll: he claims that laws of nature exist, and that whatever is referred to by the nomic concepts mentioned in Carroll's quotation, is supervenient upon which kind of laws there are.

Since in this paper we are exploring the possibility that $\Psi$ be a nomological entity, we must obviously focus on the ontological version of nomic primitivism that, unfortunately, is subject to various shortcomings, which force us either to reject it, or to endorse the view that it too is committed to the view that laws belong to the realm of the mathematical entities.

A first problem with Maudlin's ontic primitivism is the following: unlike the *conceptual* irreducibility of nomic concepts to other notions, it is not clear what it *means* to claim that laws are ontically primitive. To this objection it can be replied that if one does not want to beg the question against Maudlin's position, one ought to admit that there is a sense in which ontic nomological primitivism cannot be further understood, precisely because the notion of law is regarded as non-analyzable. However this replies raises at least two additional worries.[22]

It is true that in philosophy, as well in other disciplines, we must start from somewhere: the explanatory consequences of taking a notion $N$ as primitive legitimate the choice of $N$. However, unlike what happens in mathematics – where one often relies on axioms giving an implicit definition of whatever other concept appears in them – in philosophy, when we do not understand $N$, we are left in the dark. When a concept *N* is more obscure than a concept *P*, and we declare *N* (as well as its putative reference) "primitive", we seem to be solving a philosophical problem by *fiat*: at least in common sense and ordinary language, laws seem to be less intuitively understood than properties.

---

[22] For the following objections, see Dorato and Esfeld (2015) and Dorato and Laudisa (2015).

This first objection, however, might be tackled: intuitions about what is clear and what cries out for an explanation are vague and may vary from subject to subject. Furthermore, common sense cannot be the arbiter in metaphysical questions. A plausible reading of "ontically primitive" with respect to laws could be that there are mind-independent *global, physical facts*, which are the supervenience basis, or the ontic grounding for, properties, dispositions, or causal facts that – according to dispositionalist – are mistakenly regarded as the truth-makers of the propositions that express the "laws of science", namely those that we find in physics textbooks. This formulation, however, brings with itself the second, decisive difficulty.

Since the (approximately true) propositions regarded as truth-bearers in physics are typically differential equations, for the primitivist about laws the existence of physically necessary facts, must be contrasted with the existence of merely *contingent* facts, typically lying in hypersurfaces of simultaneity, and specifying the initial or boundary conditions to which the equations are applied. But how can the primitivist distinguish between the modally loaded, nomic facts, and the contingent facts, if both are *facts*? Clearly, ontic primitivists about laws cannot ground the distinction between nomic and contingent truths on the fact that nomic global facts hold in all *physically possible worlds*, lest laws of nature, by needing the latter notion for their specification, lose their primitivity. In other words, ontic nomic primitivism is subject to those difficulties that afflict certain versions of Humean regularism, which must distinguish between laws and mere regularities in terms of concepts (like counterfactuals) that become more fundamental than those characterizing the original empiricist position. If ontic nomic primitivism presupposed the existence of physically possible worlds, it would be refuted.

The only way out from this difficulty seems to lie in the acceptance of a mathematical approach to nomic primitivism, which, as announced, leads us back to hypothesis (ii). In fact, if we follow Maudlin in claiming that laws do not presuppose the existence of physically possible worlds, but they rather *determine* the physically possible worlds that satisfy the equations expressed by scientific laws, they can do so only in virtue of mathematical, structural features that all of these models satisfy. Independently of the fact that one can either be realist about mathematical models of physical theories or about physically possible worlds, both models and worlds, whether constructed or not, are non-physical. And only the mathematical features of laws can determine properties of other abstract entities like models or physically possible worlds. In a word, ontic nomic primitivism is either wrong or commits us to the view that laws of nature, and therefore also the universal wave function viewed as a nomological "something", are mathematical entities, which confirms the conclusion already reached in (ii).

Since modal realism about laws in any case raises the question whether it is laws *qua* mathematical entities or rather instantiated properties or dispositions that are primitive, we can discuss the second indirect way to wave function realism, namely dispositionalism that, like ontic nomic primitivism, is a modal realist view of laws, and therefore anti-Humean.

## 4 The Wave Function and Dispositionalism

We will see that dispositionalism turns out to be rather close to primitivism and therefore to the view that a realistic understanding of the wave function is committed to view that it a causally inert, non-spatiotemporally extended entity.

It has been claimed that of the main advantages of attributing properties the role of the truth-makers for laws of science is that, in particular in quantum mechanics, any problematic commitment to an abstract configuration space is avoided (see Suárez 2007, Dorato and Esfeld 2010). According to dispositionalism, the guiding law in Bohmian mechanics and dynamical reduction models *à la* GRW supervene on the fact that particles and flashes or fields respectively possess dispositions or causal powers. In a word, in the property-first view of laws applied to our case study, there is a primitive ontology of entities in spacetime that is endowed with certain properties or dispositions that manifest themselves *spontaneously* in accordance with the laws, in the sense that it is the nature of these properties that determines what the laws are.

Two provisos must be added at this point. The first is that this spontaneity depends on the fact that both in GRW-type theories and in Bohmian mechanics, the dispositions in question are not stimulated by something external. In the latter theory, for instance, the global configuration of particles is all there is, so there is nothing external to it, while in GRW-type theories the processes of localization happen at random times and are therefore not dependent on something external. For the sake of the argument, here I will dismiss the objection that dispositions must be stimulated by something external by definition: if this were true, for example, uranium could not be attributed a spontaneous disposition to decay. A partial justification for this quick rebuttal is that this criticism seems to depend on what the *meaning* of "disposition" is taken to be, and as such it is largely a matter of words. Furthermore, if powers could unfold in time spontaneously, we could treat these dispositions as powers (see also Esfeld, Lam and Hubert 2015).

The second proviso is that properties can of course be viewed in different ways, but for the sake of the argument here I will take for granted that the properties of any entity *E* are given by the causal powers or dispositions of $E$, in such a way that the nomic role of the properties are essential to $E$ (Ellis 2001, Harré and Madden 1975, Mumford and Anjum 2001). This allows me to sidestep the question whether quidditism – the view that the identity of a property of an entity $E$ does not depend on the causal powers of $E$ – is reasonable (among others, see Psillos 2006, p.18).

More in detail, in Bohmian mechanics, for instance, a disposition is attributed to the configuration of all the particles in the universe at any given time $t$ (recall that in this context we are presupposing Newtonian spacetime), which then manifests itself in the velocity of each particle at $t$. In other words, in the property-first view, the universal wave-function at that time stands for such a dispositional property as it is instantiated by the whole configuration; furthermore, in virtue of the guiding equation, the evolution of the wave-function determines how such a dispositional property manifests itself in the temporal development of the position of the particles (by fixing their velocity) (see Esfeld et al. 2013). This explains quite clearly why, in this dispositionalist approach, the kind of realism about the universal wave function is *indirect*. This means that $\Psi$ as such does not exist as a nomological entity except in the mathematical sense specified above, but also that the guiding law *refers* to a global dispositional property, which is a real property of the whole configuration of particle in the universe and is presumably causally active and therefore physical.

However, since the genuine wave function has a universal, global character, and since the dispositional property in question is instantiated by the whole configuration of particles, this disposition is instantiated by the whole universe. But then, aren't universal laws instantiated by the universe also according to nomic primitivists? If the answer to this rhetorical question is in the positive, then we have a *prima facie* argument to the effect that in the case of quantum mechanics at least, there is no real ontological difference between primitivism and dispositionalism. Given quantum holism, and given that in the former approach the wave function has a mathematical status, it should have the same status also in the former approach.

It could be objected that the dispositional property "having a certain position" with its relative manifestation is instantiated by each particle and, therefore, is instantiated at each time exactly where the particles are, but not everywhere as claimed above.[23] In this case, the dispositional property would not be global and mathematical, because it would not refer to the abstract and causally inert *set* of particles, but rather to each of the localized particles in spacetime. Here is how the objection goes: suppose a universe with only three Bohmian particles: $a$, $b$ and $c$ at a given time $t$. The velocity of particle $a$ non-locally depends on the dispositional properties of the other two particles, a property having to do with their positions. The same holds for particles $b$ and $c$. The manifestation of the properties locally instantiated by any pair of particles is the third particle's velocity, and therefore the direction and the speed of its motion. If this were the case, the global character of quantum dispositionalism in Bohmian mechanics would seem to be jeopardized and the abstract nature of the wave function would be refuted.

---

[23] I owe this objection to Albert Solé.

However, this "local" interpretation neglects an essential point that was already discussed above, namely that according to this objection the wave function, and therefore the property associated to each of the three particles, becomes relative to the three subsystems of the universe. As such, it is only effective, i.e., is used only for practical purposes, while the true objective property is the universal wave function of the universe (see above). But the universal wave function is a dispositional property of the configuration of all particles and therefore *a property of a set of all particles and not a causal property of each particle in the set*. But sets, after all, are neither causally active nor in spacetime, and so they count as non-physical, i.e., mathematical. As such, a set of particles is mathematical feature of the universe, in the same sense in which, according to ontic nomic primitivism, the mathematical models with which we represent the physical world according to quantum mechanics are fixed by the universal wave function.

Primitivism might insist in calling the global fact that there is a wave function in the universe a law valid everywhere and every-when, while dispositionalists will argue that there is universal *property* instantiated by the universe that make true the scientific law defined in the mathematical model.[24] But the main point here is that both the global nomic fact invoked by the ontic nomic primitivists and the property of the whole configuration of particles invoked by dispositionalist are – unlike the particles, which are primitive in the sense of being concrete and physical inhabitants of the spatiotemporal world – *mathematical*, which is exactly the main thesis of the paper.

This conclusion holds also for dispositionalism as it can be applied to other theories of quantum mechanics admitting a primitive ontology. In dynamical reduction models à la GRW, and more precisely in their mass density version, it is this density *as a whole* that instantiates a probabilistic dispositional property (a propensity) for spontaneous "contractions" or localizations of the field. On the other hand, in the flash-version of GRW it is the configuration of non-massless entities as a whole that instantiates a dispositional property to localize in a flash, and such localizations are powers that manifests themselves in the occurrence of later flashes. This property is represented in the mathematical model by the universal wave function as it is defined in GRW's non-linear equation, which also prescribes the probabilities for the occurrence of later flashes, in which could be grounded in time asymmetric propensities (Dorato and Esfeld 2010). Also in the case of GRW, it is no longer possible to maintain that quantum laws are grounded in local or intrinsic properties of particles.[25]

---

[24] For a sketch of argument against an identification of the two modal views of laws, which pertains to the explanatory power of dispositionalism, see Dorato and Esfeld (2015).

In a word if the holism of quantum mechanics compels dispositionalism "to go global" and if anything global, that is, spatiotemporally universal *qua* valid everywhere and at any time, is a law, the difference between primitivism about $\Psi$ and dispositionalists seems to depend only on matters of explanatory priority. It follows that according both to the view that the universal wave function is a nomological entity and to quantum dispositionalism, the wave function becomes a mathematical feature of the physical universe.

## 5 Configuration space realism as a final attempt to defend wave function realism

The discussion raised by Bell and Albert's view that the wave function is a physical field has been rather influential in the last decade, generating a debate between those who insist on the centrality of a primitive ontology and therefore of spacetime to make sense of quantum mechanics (Allori et al. 2008), and those who stress the primitive nature of configuration space, which is regarded as the main *explanans* of the world of the emergent three-dimensional objects of our experience.

Configuration or state-space realist (Albert 1996, 2013, North 2013) – and, in a different sense Everettians (Wallace 2012, 2013) – try to account for the classical world of our three-dimensional experience in terms of an ontology that regards the wave function as basic and fundamental. This fact in turn involves a belief in the existence of configuration space as the physical arena for its dynamical evolution. In this section I want to claim that until we are given a non-hand waving and precise *explanation* of the emergence of table and chairs from the ontology of configuration space realism, this proposal should not be taken seriously (see also Monton 2006).

In order to give substance to my criticism, here I will concentrate on Ney's (2013) and Wallace's (2004), as it seems representative of two important approaches to this formidable explanatory question. Ney is clearly aware of the merely promissory nature of her reductive strategy: "here we are considering an object's decomposition not into its intuitive parts but into various modes at each point in configuration space that are instantiated to varying degrees. Each mode corresponds to a slightly different classical version of itself. For example, my desk exists in the wave function ontology in virtue of having many of these different modes instantiated to a sufficient degree (amplitude) in the wave function." (Ney 2013, p. 181).

---

[25] Darby (2012) also defends a global approach to supervenience, which turns the Humean mosaic into a lawlike feature of the universe, that is, a property of the universe as such.

My first comment to this quotation stresses the vague and unclear notion of "mode": what is a "mode" and how does it relate to the explanation and reduction of objects to their wave functions? Secondly, talking of a "slightly difference classical version of itself" looks like an implicit commitment to the many worlds ontology, not clearly spelled out, and opening its flank to the difficulties of this view. But aside from this worry, that here will not be developed, the claim that the amplitude of the wave function concentrating in a region of configuration space can explain – in a way that is both reductive and non-eliminationist – the *position* of tables, chairs and human beings is neither a solution to the explanatory problem, nor "a start at an account of ontological reduction that can be used by the wave function realist" (Ney 2013, p. 181), since it is just the statement of the problem.

It is in fact not sufficient to eliminate, as Ney plausibly proposes, mereological constraints from the Putnam/Oppenheim's account of reduction: classical mereology is certainly refuted the quantum mechanics, and also by previous physical theories (Healey 2013). The problem is whether the positions or the arrangement of particles in a table – all of which are supposed to have a definite position thanks to the existence of peaks of the wave function in certain regions of configuration space – are sufficient to explain all of its macroscopic properties. At most, the positions of the particles of a table can account for the definiteness of its position as a *macroscopic* object, as in Goldstein and Zanghì's approach to Bohmian mechanics. But in the case of configuration space realism, we are not just trying to solve the measurement problem. Rather, we are forced to use *just* the definite positions of particles, regarded as an emergent property of the fundamental ontology of configuration-space quantum theory, to explain *all* the other properties of macroscopic objects.

While it might be objected that this problem is faced by all interpretations of quantum mechanics, the main trouble with the two-tiered model of reduction proposed by Ney is caused by the transitivity of the relation of explanation. One step of the reduction is to explain the emergence of the definite positions of Bohmian particles from features of the wave function evolving in configuration space. The second step is the familiar one of using the position to account for the definite position of macroscopic objects. Even if the relation of reduction were non-transitive, if we think that Bohmian mechanics is primarily about particles in spacetime, we are also allowed to take for granted the existence of macroscopic objects as a feature of the world that does *not* emerge out of a $3N$ dimensional physical space. So if we start with a primitive ontology of spatiotemporally extended or causally efficacious entity we need not assume that the position of quantum particles is sufficient to account for all its other macro-properties, because these can be regarded as ontologically self-standing and only explanatory dependent. On the contrary if we begin with a wave function ontology in configuration space, it becomes much more difficult to explain the properties of macroscopic objects.

We might ask: can this second reduction be accomplished by talking about macroscopic objects as Dennettian "patterns" of atoms? (Wallace 2004, p. 635), a position that Ney explicitly rejects? The problem here involves the unclear concept of a "pattern", which seems epistemic in nature. Dennett claims that a pattern, as a belief in a conscious being, is something that exists at an intermediate level between "industrial-strength" (full-blown) realism and fictionalism because he argues that the distinction between realism and instrumentalism should be dropped (1991, p.51), in favor of the distinction between what is useful or not useful for science.

However, note that this deflationary approach to "pattern" in the context of a debate about the (full-blown) *reality* of the universal wave function is misleading on two counts. The first is that the universal wave function would also become something intermediate between full-blown realism and fictionalism, something that Wallace and co. should oppose, given that according to them the wave function is the fundamental entity of the physical world. Of course, one could use this deflationary position only for the world of our experience. But here lies the second problem of Wallace's analysis: the world of the manifest image, *qua* pattern in the universal wave function, would also become something intermediate between reality and fiction, a position that would leave most of us, qua common sense realists who did not buy into Berkeley's philosophy, rather unsatisfied!

The fact that the notion of pattern has a mere epistemic dimension is suggested by the following quotation: "a pattern is "by definition" a candidate for pattern recognition. (It is this loose but unbreakable link to observers or perspectives, of course, that makes "pattern" an attractive term to someone perched between instrumentalism and industrial-strength realism.)" (Dennett 1991, p. 32). The idea is therefore that: "A pattern exists in some data – is real – if there is a description of the data that is more efficient than the bit map, whether or not anyone can concoct it" (ibid. p.34). Once again, this view makes the reality of the world of our experience relative to the choice of the language that yields more informativity and predictive strength. Even if this language has not been concocted so far.

The configuration space realist could reply that this approach to the emergence of the world of our experience is compatible with the claim that the wave function is real, even if abstract. In fact, it is a "good" abstract object, given that it helps us to predict important features of the physical world around us. And by applying Dennett's view to our case, one could be a realist also about the manifest image, since the latter could be regarded as a pattern in configuration space, even if at the moment we are not able to specify a reduction of the objects of the three dimensional worlds in terms of the wave function and the motion of particles that it purports to describe.

Unfortunately, Wallace's account is saddled with an additional problem, given by his vague and merely verbal accounts of the emergence of a three-dimensional world in terms of structural realism. His proposal is to use structural realism in order to account for the relationship between the states of classical system and quantum states as they are represented by wave packets: "wave packets states represent classical objects in motion because their dynamical behavior instantiates the dynamical behavior of those classical objects (Wallace 2013, p. 213)". So far we have been given a reasonable explanatory account of the *dynamics* of classical objects in terms of Schrödinger's equation. But then we are told: "in other words, the *structural features* of the classical world are in fact represented – to a high degree of accuracy – by structural features of the dynamics of the wave packets states" (ibid., my emphasis). In this quotation we inadvertently pass from the dynamical features of classical objects to their structural features. But unless "the structural features of the classical objects" reduce in a mysterious way to their dynamical evolution in classical spacetime, we don't know what such structural features are: are all properties of tables and chairs merely structural?[26] And what does it mean to claim that a table has only structural properties isomorphic to the dynamical features of wave packets?

In Dennett's original proposal (1991), the existence of macroscopic objects is predicated on the usefulness of this assumption for predictive and explanatory roles. In other words, a real pattern is inferred as the best explanation of what happens at the relevant higher-level ontic level; on the other hand such macroscopic patterns are our best epistemic warrant to infer the existence of entities at a more fundamental level. Even granting that this view is not problematic or epistemically circular, the *dynamical* features of classical objects are but a minimal part of their properties in three-dimensional space.

**6 Conclusions**

In order to try to come a more conciliatory position between the three alternative interpretations of the wave function that we have been discussing under (i) (ii) and (iii), one might try to argue that, as far as the ontic status of the wave function is concerned, the difference between Goldstein & Zanghì and dispositionalists on the one hand, and Albert or Ney (or possibly North 2013) on the other, does not involve the *ontology* of quantum theory but rather only the *epistemology* of the wave function. That is, it involves just explanatory priority. After all, both groups regard $\Psi$ as existent, but while the former group claims that $\Psi$ is an *abstract* physical entity, the latter affirm that it is a mathematical entity. Therefore, one might want to continue, the difference between philosophers insisting on the ontic primacy of particles or flashes and configuration space realists is in part ontological but more significantly epistemic, in so far as it can be traced to the explanatory order in which one wants to understand the physical world.

---

[26] See also Wallace (2013, pp.48-52).

I have shown in the previous section that the difference between nomological realists and dispositionalists about $\Psi$ tends to vanish when one pays attention to the fact that, due to quantum holism, both must consider mathematical entities as ontologically central. Could one argue that the same conciliatory conclusion applies even to configuration space realist and defenders of a primitive ontology made of spatiotemporally extended?

It could be rebutted that establishing what is ontologically prior ought to precede the explanatory task, so that deciding on the different ontologies is prior to deciding about which explanatory order to embrace in order to explain the world of our experience. But the lack of a reasonable explanation of the emergence of a three-dimensional world from the wave function regarded as a physical field renders this claim of the priority of ontology over epistemology only a sterile slogan. Given the underdetermination of ontological posits by physical theories, it is only the explanatory primacy of configuration space realism that could justify Albert and Ney's ontological assumptions. Therefore, until a convincing and non-hand-waving explanation of this kind is provided, betting on the mathematical, non-physical nature of the wave function – as implied by primitivism and dispositionalism – is the most reasonable option.

This conclusion will not make nominalists happy and if we had reasons to endorse nominalism in physics, we would be forced to accept a sophisticated instrumentalism about the wave function. However, despite the connection between the main question raised in this paper and nominalism, it would be unreasonable to suspend our judgment about the metaphysics of the wave function and conclude against its abstract nature until complicated and age-old questions about both the nature of properties and Platonism are settled.